\documentclass[useAMS,usenatbib,usegraphicx]{mn2e}
\title[Cosmic-Ray Heating of Cool Cores] {Stability Analysis for
Cosmic-Ray Heating of Cool Cores in Galaxy Clusters} \author[Y. Fujita,
et al.]{Yutaka Fujita$^{1}$\thanks{E-mail:
fujita@vega.ess.sci.osaka-u.ac.jp}, Sota Kimura$^{1}$, and Yutaka
Ohira$^{2}$\\ $^{1}$Department of Earth and Space Science, Graduate
School of Science, Osaka University, 1-1 Machikaneyama-cho, \\ Toyonaka,
Osaka 560-0043, Japan\\ $^{2}$Department of Physics and Mathematics,
Aoyama Gakuin University, Fuchinobe, Chuou-ku, Sagamihara 252-5258,
Japan}
\begin{document}

\date{Accepted 0000 December 15. Received 0000 December 14; in original form 0000 October 11}

\pagerange{000--000} \pubyear{0000}

\maketitle

\label{firstpage}

\begin{abstract}
 We study the heating of the cool cores in galaxy clusters by
 cosmic-rays (CRs) accelerated by the central active galactic nuclei
 (AGNs). We especially focus on the stability of the heating. The CRs
 stream with Alfv\'en waves in the intracluster medium (ICM) and heat
 the ICM. First, assuming that the heating and radiative cooling is
 balanced, we search steady state solutions for the ICM and CR profiles
 of clusters by solving a boundary value problem. The boundary
 conditions are set so that the solutions are consistent with
 observations of clusters. We find steady state solutions if the
 magnetic fields are strong enough and the association between the
 magnetic fields and the ICM is relatively weak. Then, we analyse the
 stability of the solutions via a Lagrangian perturbation analysis and
 find that the solutions are globally stable. We confirm the results by
 numerical simulations. Using the steady state solutions as the initial
 conditions, we follow the evolution of the profiles for 100~Gyr. We
 find that the profiles do not evolve on time scales much larger than
 cluster lifetimes. These results, as well as consistency with
 observations of radio mini-halos, suggest that the CR heating is a
 promising mechanism to solve the so-called ``cooling flow problem''.
\end{abstract}


\section{Introduction}

The radiative cooling time of the intracluster medium (ICM) in the cores
of galaxy clusters is often smaller than the age of the clusters
\citep{sar86a}. If there were no heating sources, massive cooling flows
toward the cluster centres would develop in the cores
\citep{fab94}. However, X-ray observations have revealed that such flows
do not exist in the cores
\citep[e.g.][]{ike97,mak01,pet01,tam01,kaa01,mat02}. This means that
some unknown heating sources prevent such flows from developing (cooling
flow problem).

AGNs at the cluster centres are often thought to be the heating
sources. However, the energy transfer mechanism from the AGNs to the
surrounding ICM is not understood. Conventional mechanical heating
(e.g. shocks, or sound waves) may cause thermal instability
\citep*{fuj05,mat06}. Thus, strong turbulence would be required to
stabilise the heating, if the cores are mainly mechanically heated.

Cosmic-rays (CRs) accelerated by the AGNs are another promising carrier
of the energy to the ICM
\citep*[e.g.][]{tuc83,rep87,rep95,col04,pfr07,jub08}.  CR streaming is
one way to deposit their energy into the ICM
\citep*{rep79,boh88,loe91,guo08a,pfr13a}. Each CR particle moves with a
velocity close to the light velocity $c$. However, in a cluster, they
are scattered by Alfv\'en waves in the ICM. Thus, the CRs effectively
move along with the waves with a bulk (streaming) velocity $v_{\rm st}$,
which is much smaller than $c$. Since the CR pressure is high around the
central AGN and waves are excited in the direction of the CR pressure
gradient \citep[e.g.][]{lon94}, the waves and the CRs scattered by them
move outward in the cluster in general. As a result, the CR pressure
does $PdV$ work against the ICM, which ultimately heats the ICM. Using
numerical simulations, we showed that the ICM can be stably heated by
the CR streaming \citep[][hereafter Paper~I]{fuj11b}. The main reason of
the stability is that the CRs stream in the ICM and the heating is not
localised around the AGN. In successive studies, we calculated
non-thermal emissions from the CRs \citep[][hereafter Paper~II]{fuj12a},
and we showed that the radial profiles of radio mini-halos observed in
clusters can be reproduced in our model \citep[][hereafter
Paper~III]{fuj13a}.

In this paper, we study the reason of the stability of the CR heating in
more detail based on a perturbation analysis and numerical
simulations. This paper is organised as follows. In \S~\ref{sec:model},
we explain our models and show the basic equations. In \S~\ref{sec:vA},
we solve those equations and obtain steady state solutions when the CR
streaming velocity is the Alfv\'en velocity. Moreover, we analyse the
stability of the steady state solutions.  In \S~\ref{sec:dis}, we
discuss the evolution of the ICM with time and radio mini-halos observed
around cluster cores. \S~\ref{sec:con} is devoted to conclusions.
Throughout this paper we assume a $\Lambda$CDM cosmology with
$\Omega_m=0.3$, $\Omega_\Lambda=0.7$, and $h=0.7$, where $H_0=100\:
h\rm\: km\: s^{-1}\: Mpc^{-1}$. We consider protons as CRs unless
otherwise mentioned.

\section{Models}
\label{sec:model}

\subsection{Equations}

The model we adopt in this study is basically the same as that in
Papers~I--III. Assuming that the cluster is spherically symmetric, the
flow equations are
\begin{equation}
\label{eq:cont}
 \frac{\partial \rho}{\partial t} 
+ \frac{1}{r^2}\frac{\partial}{\partial r}(r^2\rho u) = 0\:,
\end{equation}
\begin{equation}
\label{eq:mom}
\frac{\partial (\rho u)}{\partial t} 
+ \frac{1}{r^2}\frac{\partial}{\partial r}(r^2\rho u^2)
= \rho g-\frac{\partial}{\partial r}
(P_g + P_c + P_B)\:,
\end{equation}
\begin{eqnarray}
\label{eq:eg}
 \lefteqn{\frac{\partial e_g}{\partial t}  
+ \frac{1}{r^2}\frac{\partial}{\partial r}(r^2 u e_g)
=  -P_g \frac{1}{r^2}\frac{\partial}{\partial r}(r^2 u)}\nonumber\\
&& + \frac{1}{r^2}\frac{\partial}{\partial r} 
\left[r^2\kappa(T)\frac{\partial T}{\partial r}\right]
- n_e^2\Lambda(T) + H_{\rm st} + H_{\rm coll}\:,
\end{eqnarray}
\begin{eqnarray}
\label{eq:ec}
 \lefteqn{\frac{\partial e_c}{\partial t}  
+ \frac{1}{r^2}\frac{\partial}{\partial r}(r^2 \tilde{u} e_c)
= -P_c \frac{1}{r^2}\frac{\partial}{\partial r}(r^2 \tilde{u})}
\nonumber\\
&& + \frac{1}{r^2}\frac{\partial}{\partial r}
\left[r^2 D(\rho)\frac{\partial e_c}{\partial r}\right] 
- \Gamma_{\rm loss}
+ \dot{S}_c \:,
\end{eqnarray}
where $\rho$ is the gas density, $u$ is the gas velocity, $P_g$ is the
gas pressure, $P_c$ is the CR pressure, $P_B$ is the magnetic pressure,
$g$ is the gravitational acceleration, $\kappa(T)$ is the coefficient
for thermal conduction, $T$ is the temperature, $n_e$ is the electron
density, $\Lambda$ is the cooling function, $H_{\rm st}$ is the heating
by CR streaming, $H_{\rm coll}$ is the heating by Coulomb and hadronic
collisions, $\tilde{u}$ is the CR transport velocity, $D(\rho)$ is the
diffusion coefficient for CRs averaged over the CR spectrum,
$\Gamma_{\rm loss}$ is the energy loss by Coulomb and hadronic
collisions, and $\dot{S}_c$ is the source term of CRs. Energy densities
of the gas and the CRs are respectively defined as
$e_g=P_g/(\gamma_g-1)$ and $e_c=P_c/(\gamma_c-1)$, where
$\gamma_g(=5/3)$ and $\gamma_c$ are the adiabatic indices for the ICM
and CRs, respectively. For the latter, we assume $\gamma_c=5/3$, because
we found that the energy spectrum of the CRs must be steep and most of
the CRs have low energies by comparing model predictions with radio
observations (Papers~II and~III).

Since we later perform a perturbation analysis, we need to make the
model simpler. We do not include thermal conduction and CR diffusion
($\kappa=0$ and $D=0$). For the same reason, we assume a simple cooling
function:
\begin{eqnarray}
\lefteqn{n_e^2\Lambda(T) = \rho {\cal L}}\nonumber\\
&& = 2.1\times 10^{-27}
 \left(\frac{n_e}{\rm cm^{-3}}\right)^2
\left(\frac{T}{\rm K}\right)^{1/2}
\rm erg\: cm^{-3}\: s^{-1}
\end{eqnarray}
\citep{ryb79,kim03a,guo08b}. We also ignore the gradient of magnetic
pressure ($\partial P_B/\partial r=0$) in equation~(\ref{eq:mom}),
because it is not dynamically important.

\subsection{AGN}

In our model, CRs are accelerated by the central AGN. The source term
of the CRs is given by
\begin{eqnarray}
\lefteqn{\label{eq:dotSc}
 \dot{S}_c = \frac{3-\nu}{4\pi}
\frac{L_{\rm AGN}}{r_1^3 (r_1/r_0)^{-\nu}-r_0^3}
\left(\frac{r}{r_0}\right)^{-\nu}}\nonumber\\
&&\times (1-e^{-(r/r_0)^2})e^{-(r/r_1)^2}\:,
\end{eqnarray}
where $L_{\rm AGN}$ is the energy injection rate from the AGN. We assume
that 
\begin{eqnarray}
\label{eq:LAGN}
\lefteqn{L_{\rm AGN} = -\epsilon \dot{M} c^2} \nonumber\\
&& = 1.4\times 10^{44}{\rm erg\: s^{-1}}
\left(\frac{\epsilon}{2.5\times
     10^{-4}}\right)
\left(\frac{-\dot{M}}{10\: M_\odot\rm\: yr^{-1}}\right)
\:,
\end{eqnarray}
where $\epsilon$ is the parameter, and $\dot{M}$ is the flow rate of the
gas. In this model, accelerated CRs are first carried by buoyant bubbles
from the AGN out to large distances \citep{guo08a}. As the bubbles
filled with the CRs adiabatically rise, the CRs may escape from the
bubbles into the ICM or they may be injected into the ICM through the
shredding of the bubbles by Rayleigh-Taylor and Kelvin-Helmholtz
instabilities. Thus, the CRs are gradually injected into the ICM as the
buoyant bubbles rise. Unless otherwise mentioned, we fixed the
parameters at the values similar to the ones used in Papers~I--III
($\epsilon=2.5\times 10^{-4}$, $\nu=3.2$, $r_0=20$~kpc, and
$r_1=150$~kpc), because they give results that are consistent with
observations (see later). In Papers~I and II, we assumed that the CRs
are accelerated at the forward shock of a cocoon created through the AGN
activities. However, it may be more appropriate to assume that CR
protons are accelerated around the central black hole and that AGN jets
or winds consist of the CR protons, especially when $\epsilon$ is large
(\S~3.1 of Paper~III; see also \citealt{sik05a,tom12a}).

\subsection{Cosmic-rays}

The CR transport velocity is given by $\tilde{u}=u+v_{\rm st}$ in
equation~(\ref{eq:ec}). The most simple idea is that $v_{\rm st}$ is the
Alfv\'en velocity $v_A$, because the CRs are scattered by the Alfv\'en
waves and move with them. However, there have been debates about this
issue. It has been indicated that $v_{\rm st}$ may be much larger than
$v_A$, because in hot ICM, Alfv\'en waves may damp at small wave lengths
via interactions with thermal protons. In this case, the sound velocity
of the gas $c_s$ may be more plausible as the streaming velocity
\citep{hol79,ens11}. However, there also have been indications that the
deficit of the scattering at short wave lengths is overcome by other
effects such as mirror scattering or wave cascading
\citep[e.g.][]{fel01a,sch02,wie13a}. If this is the case, the assumption
of $v_{\rm st}=v_A$ is appropriate. We consider the latter case ($v_{\rm
st}=v_A$) in the following sections. We discuss the former case ($v_{\rm
st}=c_s$) in Appendix.

In this study, we investigate steady state solutions. Therefore, we
assume that the growth of the waves is balanced with non-linear wave
damping. In other words, the CR energy consumed to grow the waves is
equal to the wave energy that is put into the ICM through the wave
damping. Thus, the heating term of the CR streaming is given by
\begin{equation}
H_{\rm st}=-v_{\rm st}\frac{\partial P_c}{\partial r}\:.
\end{equation}
Note that $v_{\rm st}>0$ and $\partial P_c/\partial r<0$ in our
calculations.

In Papers~I and~II, we studied the evolution of wave amplification. In
those studies, we conservatively assumed that the balance between the
wave growth and the damping is archived when the wave energy density
reaches to that of the background magnetic fields, because some of
non-linear damping mechanisms should be effective after that. As shown
in Paper~I (its Fig.~3), the time scale in which the growth and the
damping is balanced is $\sim$~Gyr. Since it is less than the time scale
of a cluster lifetime ($\sim 10$~Gyr), the assumption of the balance in
this study is justified. Moreover, if the damping is very efficient, the
saturation may be achieved even faster. For example, if the time scale
of the damping is proportional to the inverse of the gyro frequency of
CR particles (e.g. non-linear Landau damping; \citealt{fel01a}), the
saturated wave energy density is much smaller than that of the
background magnetic fields.

We ignore Coulomb and hadronic collisions ($H_{\rm coll}$=0 and
$\Gamma_{\rm loss}=0$) in equations~(\ref{eq:eg}) and~(\ref{eq:ec}),
because they do not much affect the results as follows (see also
Papers~I and~III). The collisional heating term is given by $H_{\rm
coll}=-\Gamma_c-\Gamma_h/6$, and the loss term is given by $\Gamma_{\rm
loss}=-\Gamma_c-\Gamma_h$, where $\Gamma_c$ is the Coulomb loss rate and
$\Gamma_h$ is the hadronic loss rate (Paper~III). In Paper~III, we
estimated that
\begin{equation}
\label{eq:gamc}
 \Gamma_c
=\chi_c
\left(\frac{n_e}{\rm cm^{-3}}\right)
\left(\frac{e_c}{\rm erg\: cm^{-3}}\right)
\rm\: erg\: s^{-1}\: cm^{-3} \:,
\end{equation}
\begin{equation}
\label{eq:gamh}
 \Gamma_h=\chi_h
\left(\frac{n_e}{\rm cm^{-3}}\right)
\left(\frac{e_c}{\rm erg\: cm^{-3}}\right)
\rm\: erg\: s^{-1}\: cm^{-3} \:,
\end{equation}
where $\chi_c=-7.3\times 10^{-16}$ and $\chi_h=-1.5\times 10^{-17}$.  In
these estimations, we assumed that the CR momentum spectrum is given by
a power low ($\propto p^x$), and the index is $x=3$. We chose $x=3$
because it is consistent with observations of radio mini-halos in
clusters (Paper~III). Moreover, we assumed that the minimum momentum is
$p_{\rm min}c=137$~MeV, at which the effect of Coulomb collision is
maximum (Paper~III). For these values, we have confirmed that $H_{\rm
coll}$ and $\Gamma_{\rm loss}$ can be ignored for the results in the
following sections. We note that the index could be as large as $x=3.5$
to be consistent with the observations when $v_{\rm st}=v_A$
(Paper~III). We estimated $\Gamma_c/(n_e e_c)$ and $\Gamma_h/(n_e e_c)$
for $x=3.5$ and $p_{\rm min}c=137$~MeV, and found that
$\chi_c=-7.6\times 10^{-16}$ and $\chi_h=-4.9\times 10^{-18}$. Thus,
$H_{\rm coll}/(n_e e_c)$ and $\Gamma_{\rm loss}/(n_e e_c)$ are not much
different from those when $x=3$ and $p_{\rm min}c=137$~MeV.

We also changed $p_{\rm min}$ for $x=3$. When $p_{\rm min}c=43$~MeV,
$\chi_c=-2.7\times 10^{-16}$ and $\chi_h=-1.6\times 10^{-18}$, and when
$p_{\rm min}c=440$~MeV, $\chi_c=-3.8\times 10^{-16}$ and
$\chi_h=-1.2\times 10^{-16}$. These mean that $H_{\rm coll}/(n_e e_c)$
and $\Gamma_{\rm loss}/(n_e e_c)$ are smaller than those when $p_{\rm
min}c=137$~MeV. Thus, we can ignore $H_{\rm coll}$ and $\Gamma_{\rm
loss}$. It is to be noted that when $x>3$, one can obtain $\Gamma_c/(n_c
e_c)\propto p_{\rm min}^2$ regardless of $x$ in the limit of small
$p_{\rm min}$ (see equations~[7] and [8] in Paper~III).

\subsection{Cluster}

For the gravitational mass profile, we adopt the NFW model
\citep*{nav97}, although there is a debate about the slope of the
central cusp \citep[e.g.][]{fuk97}. For the NFW profile, the mass
distribution is written as
\begin{equation}
\label{eq:Mr}
M(r)=M_{200}\frac{\ln(1+r/r_s)-r/r_s/(1+r/r_s)}
{\ln(1+c_{200})-c_{200}/(1+c_{200})}\:.
\end{equation}
For this equation, we define $r_q$ as the cluster radius within which
the average mass density is $q$ times the critical density of the
universe $\rho_{\rm cr}(z)$ at redshift $z$. Moreover, $M_q$ is the mass
of the cluster within $r_q$, $r_s$ is the characteristic radius, and
$c_q=r_q/r_s$. In equation~(\ref{eq:Mr}), we set $q=200$. From the
definition, we have
\begin{equation}
\label{eq:rd}
r_q=\left[\frac{3 M_q}{4\pi q\rho_{\rm cr}(z)}\right]
^{1/3}
\:.
\end{equation}

The mass $M_{200}$ is derived from the ICM temperature outside the core,
$T_{\rm out}$. Based on a statistical study, \citet{che07a} obtained a
scaling relation of
\begin{equation}
\label{eq:M500}
M_{500} = 2.6\times 10^{14} h^{-1}\left(\frac{T_{\rm out}}
{4\rm\: keV}\right)^{1.48}\: M_\odot\:.
\end{equation}
If we assume the NFW profile, $M_{500}$ can be converted to $M_{200}$
using a relation of $M_q\propto q^{-0.266}$ \citep*{hor99a}.
Theoretically, $c_{200}$ is expected to be a weakly decreasing function
of $M_{200}$ \citep[e.g.][]{duf08a}. However, observations have shown
that it has a very large scatter \citep{oka10a,ett10a}, which may
reflect a broad range of the formation epochs of clusters with a give
mass \citep[e.g.][]{fuj99d}. Thus, we fix it at $c_{200}=5$. Using
equations~(\ref{eq:Mr})--(\ref{eq:M500}), we can
determine $M(r)$ of a cluster with given $T_{\rm out}$ and $z$. The
gravitational acceleration for the NFW profile is $g_{\rm
NFW}(r)=-GM(r)/r^2$.

In addition to $g_{\rm NFW}$, we include the gravitation from the
central cD galaxy. We use the one obtained by \citet*{mat06} and used in
Papers~I-III:
\begin{equation}
g_{\rm cD}(r)
= \left[\left(\frac{r^{0.5975}}{3.206\times 10^{-7}}\right)^s
+ \left(\frac{r^{1.849}}{1.861\times 10^{-6}}\right)^s
\right]^{-1/s}
\end{equation}
in cgs with $s=0.9$ and $r$ in kpc. We assume that $g_{\rm cD}$ does not
depend on host clusters. Thus, the total gravitational acceleration in a
cluster is $g=g_{\rm NFW}+g_{\rm cD}$. The inclusion of $g_{\rm cD}$
does not qualitatively change the results.

\section{Results}
\label{sec:vA}

\subsection{Steady State Solutions}
\label{sec:steady_vA}

First, we derive steady state solutions ($\partial/\partial t=0$). The
procedure is basically the same as that in previous studies
\citep{kim03a,guo08b}. From equation~(\ref{eq:cont}), the mass flow rate
is given by $\dot{M}=4\pi r^2 \rho u$. Other three
equations~(\ref{eq:mom})--(\ref{eq:ec}) can be rewritten as
\begin{equation}
\label{eq:mom2}
\frac{1}{r^2}\frac{d}{d r}(r^2\rho u^2)
= - \rho \frac{G M(r)}{r^2}-\frac{d}{d r}
(P_g + P_c)\:,
\end{equation}
\begin{eqnarray}
\label{eq:eg2}
\frac{1}{r^2}\frac{d}{d r}(r^2 u e_g)
= -P_g \frac{1}{r^2}\frac{d}{d r}(r^2 u) 
- n_e^2\Lambda(T) + H_{\rm st}\:,
\end{eqnarray}
\begin{equation}
\label{eq:ec2}
\frac{1}{r^2}\frac{d}{d r}(r^2 \tilde{u} e_c)
= -P_c \frac{1}{r^2}\frac{d}{d r}(r^2 \tilde{u}) 
+ \dot{S}_c \:.
\end{equation}
We solve these ordinary differential equations for $r_{\rm in}<r<r_{\rm
out}$, where $r_{\rm in}=3$~kpc and $r_{\rm out}=1$~Mpc using {\it
Mathematica~9} \footnote{http://www.wolfram.com/}. They can be solved as
a boundary value problem. We impose the following boundary conditions:
\begin{equation}
\label{eq:bc_n0}
n_e(r_{\rm in})=n_0\:,
\end{equation}
\begin{equation}
T(r_{\rm in}) = T_{\rm in}\:,
\end{equation}
\begin{equation}
T(r_{\rm out}) = T_{\rm out}\:,
\end{equation}
\begin{equation}
\label{eq:bc_Pc0}
P_c(r_{\rm in}) = P_{c0}\:.
\end{equation}
Equations~(\ref{eq:mom2})--(\ref{eq:ec2}) form an eigenvalue problem in
which $\dot{M}$ is the eigenvalue.

When $v_{\rm st}=v_A=B/\sqrt{4\pi\rho}$, we need to specify magnetic
fields $B$ in the ICM. We assume that $B=B_0\: (n_e/0.016{\rm\:
cm^{-3}})^b$. We could not often find steady state solutions for too
small $B_0$ and/or too large $b$. Thus, we assume that $B_0=1.0\times
10^{-5}\: \mu\rm G$ and $b=0.4$, although we assumed that $B_0=1.0\times
10^{-5}\: \mu\rm G$ and $b=2/3$ in Paper~III. The fairly small value of
$b$ may mean that the coupling between magnetic fields and the ICM is
weak, which may be realised when the magnetic fields are rather radially
extended in a cluster. The small $b$ is also favourable to suppress the
development of local instabilities (see \S~\ref{sec:num_vA}).

\begin{figure}
\includegraphics[width=84mm]{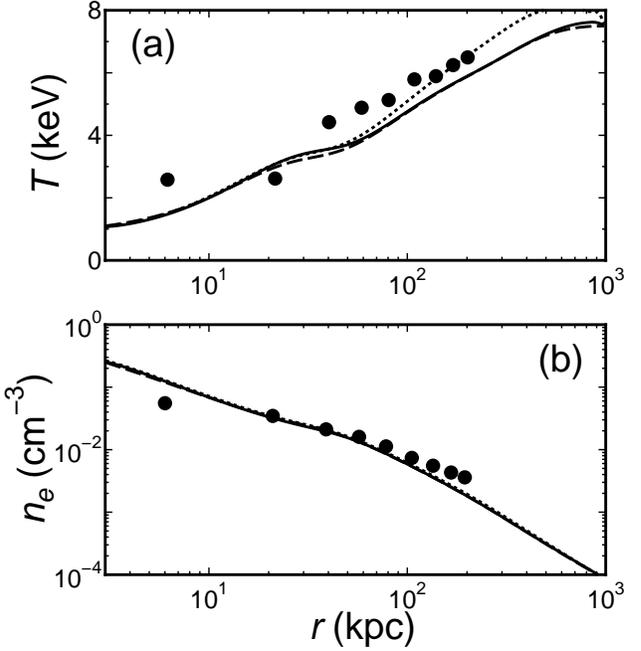} \caption{(a) Temperature and (b)
density profiles for A1795. Dashed lines show the steady state solution
or the initial profiles for the numerical simulation.  Solid and Dotted
lines are the results of numerical simulation at $t=40$~Gyr and
$100$~Gyr, respectively. Filled circles represent observations
\citep{ett02a}. Error bars are omitted.}  \label{fig:Tn_a1795_vA}
\end{figure}

\begin{figure}
\includegraphics[width=84mm]{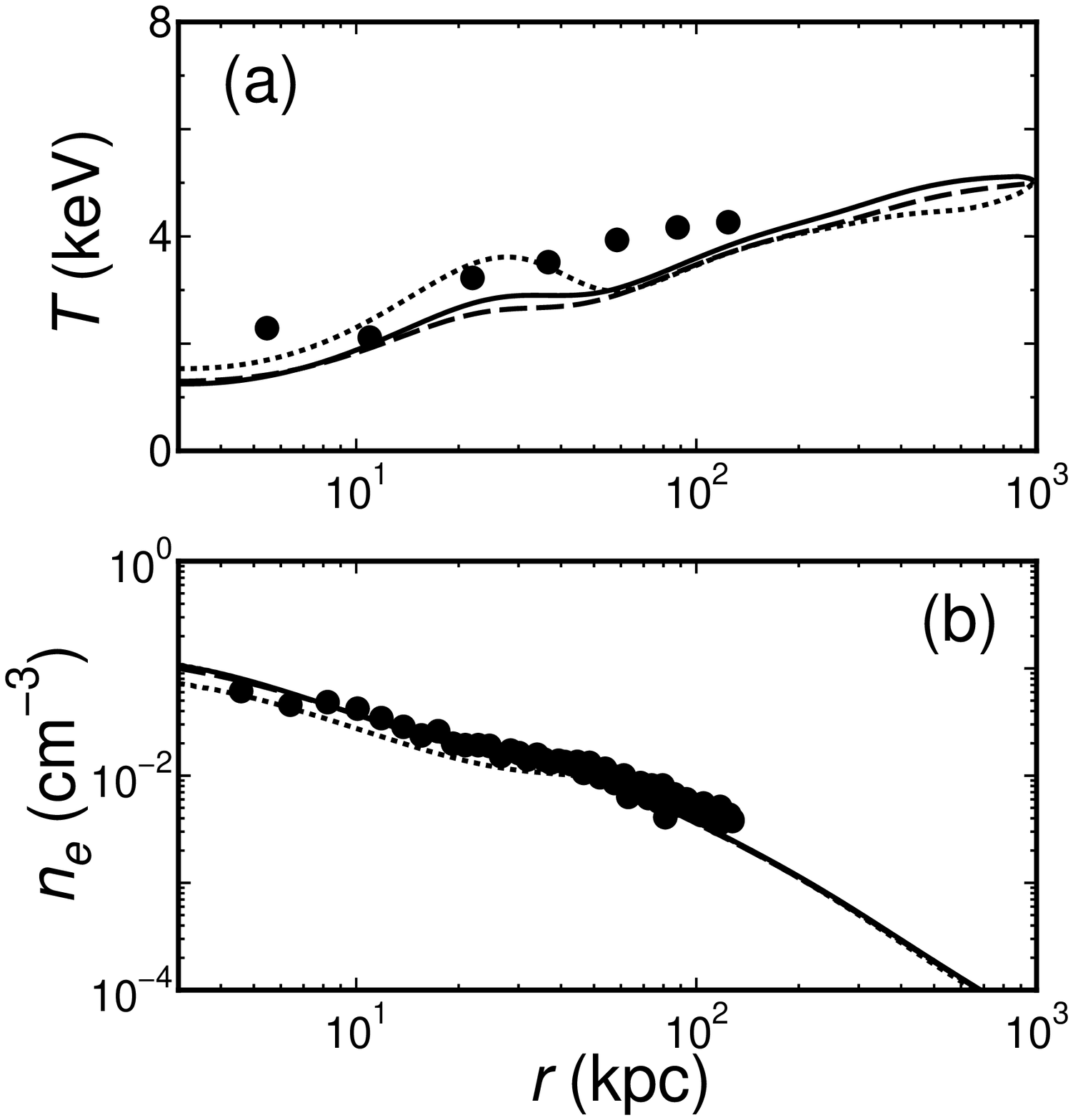} \caption{Same as
Fig.~\ref{fig:Tn_a1795_vA} but for A2199. Filled circles represent
observations \citep{jon02a}.}
\label{fig:Tn_a2199_vA}
\end{figure}

\begin{figure}
\includegraphics[width=84mm]{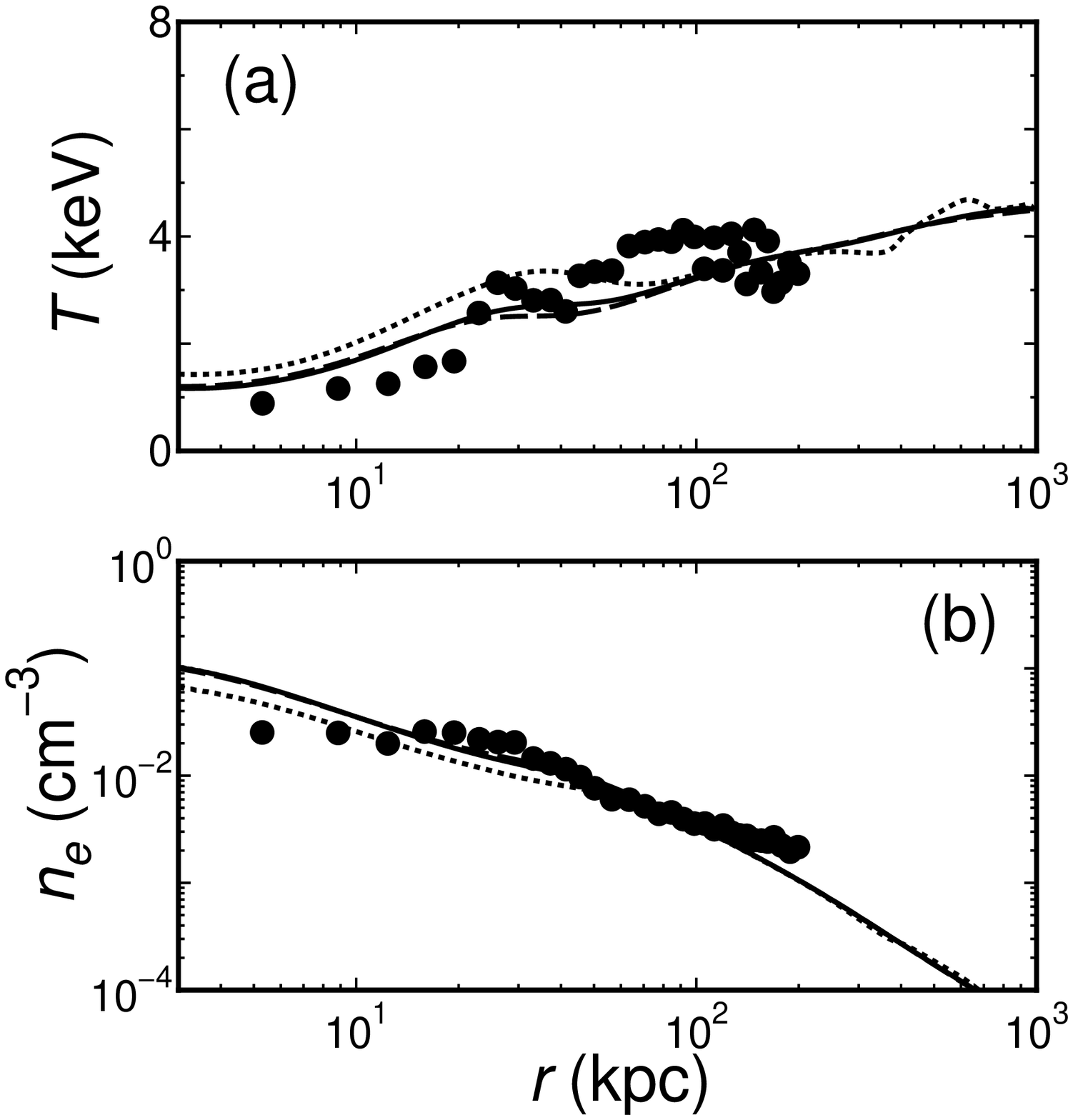} \caption{Same as
Fig.~\ref{fig:Tn_a1795_vA} but for A2052. Filled circles represent
observations \citep{bla11a}.}
\label{fig:Tn_a2052_vA}
\end{figure}

\begin{figure}
\includegraphics[width=84mm]{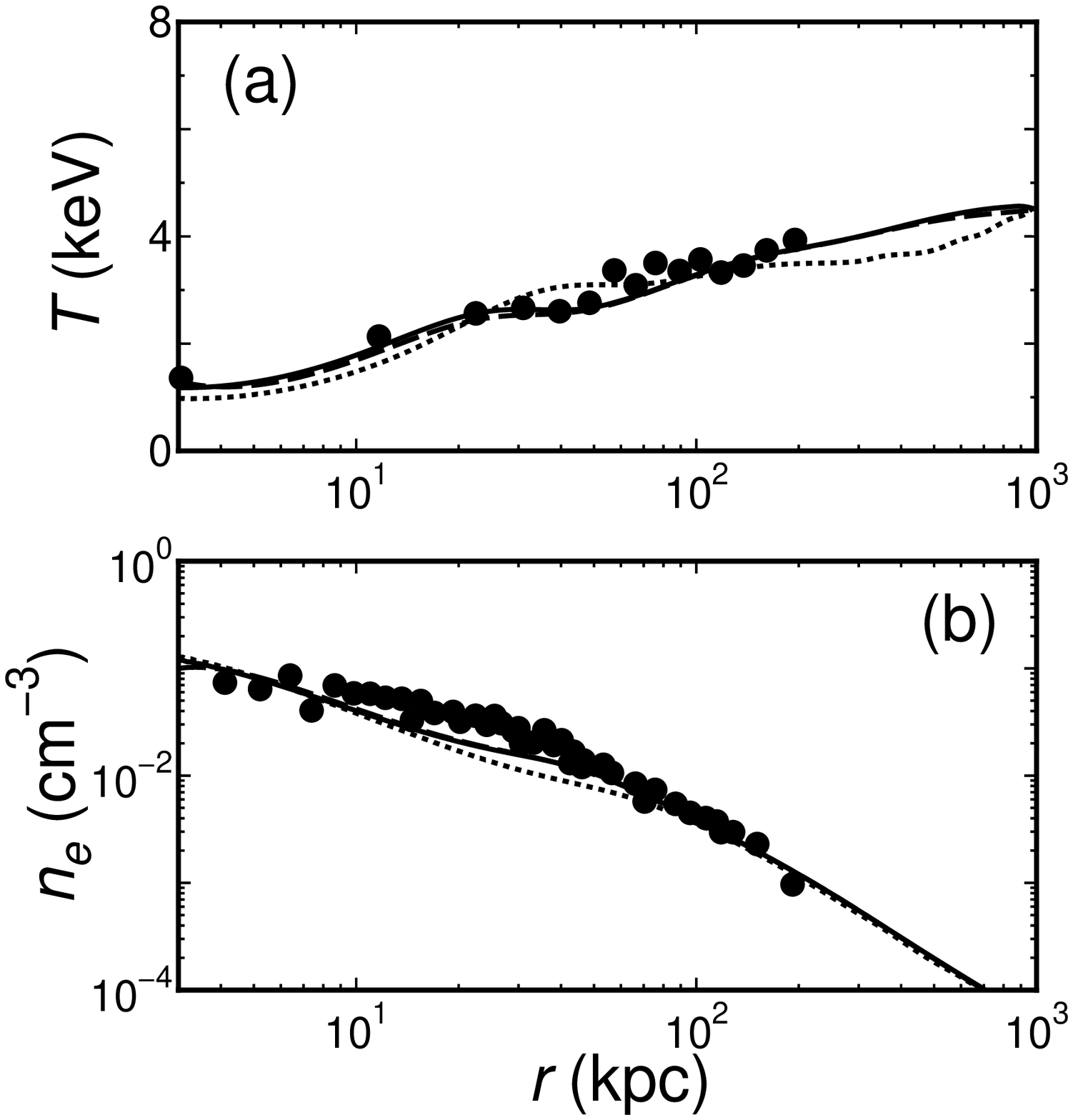} \caption{Same as
Fig.~\ref{fig:Tn_a1795_vA} but for A2597. Filled circles represent
observations \citep{mcn01a}.}
\label{fig:Tn_a2597_vA}
\end{figure}

We construct steady state solutions for four clusters with various
temperatures (A1795, A2199, A2052, and A2597), which were studied by
\citet{guo08b}. In Table~\ref{tab:cl}, we give the boundary values
$n_0$, $T_{\rm in}$, and $T_{\rm out}$ when $v_{\rm st}=v_A$, which were
chosen to be almost consistent with observations
(Figs.~\ref{fig:Tn_a1795_vA}--\ref{fig:Tn_a2597_vA}). Although there are
no direct observations of $P_{c0}$, the values of $P_{\rm c0}/P_{g0}$,
where $P_{g0}=P_{g}(r_{\rm in})$, are chosen so that $dP_{\rm c0}/dr$ is
not positive and close to zero at $r=r_{\rm in}$.

Dashed lines in Figs.~\ref{fig:Tn_a1795_vA}--\ref{fig:Tn_a2597_vA} show
the steady state solutions we obtained. Derived mass flow rates,
$\dot{M}$, are shown in Table~\ref{tab:cl}. It is to be noted that
$\dot{M}$ is the mass flow that passes the inner boundary ($r_{\rm
in}=3$~kpc), and that not all the mass needs to fall into the central
black hole. Most of the gas may become cold gas or be consumed by star
formation in the central galaxy \citep[e.g.][]{bre06b,mcn12a}. For
comparison, we show observations of the four clusters in the
figures. The steady state solutions can generally reproduce the
observations. Since our model is rather simple (we do not include
additional mechanical heating, for example), we think that it would be
useless to perfectly fit the solutions with the observations.

\begin{figure}
\includegraphics[width=84mm]{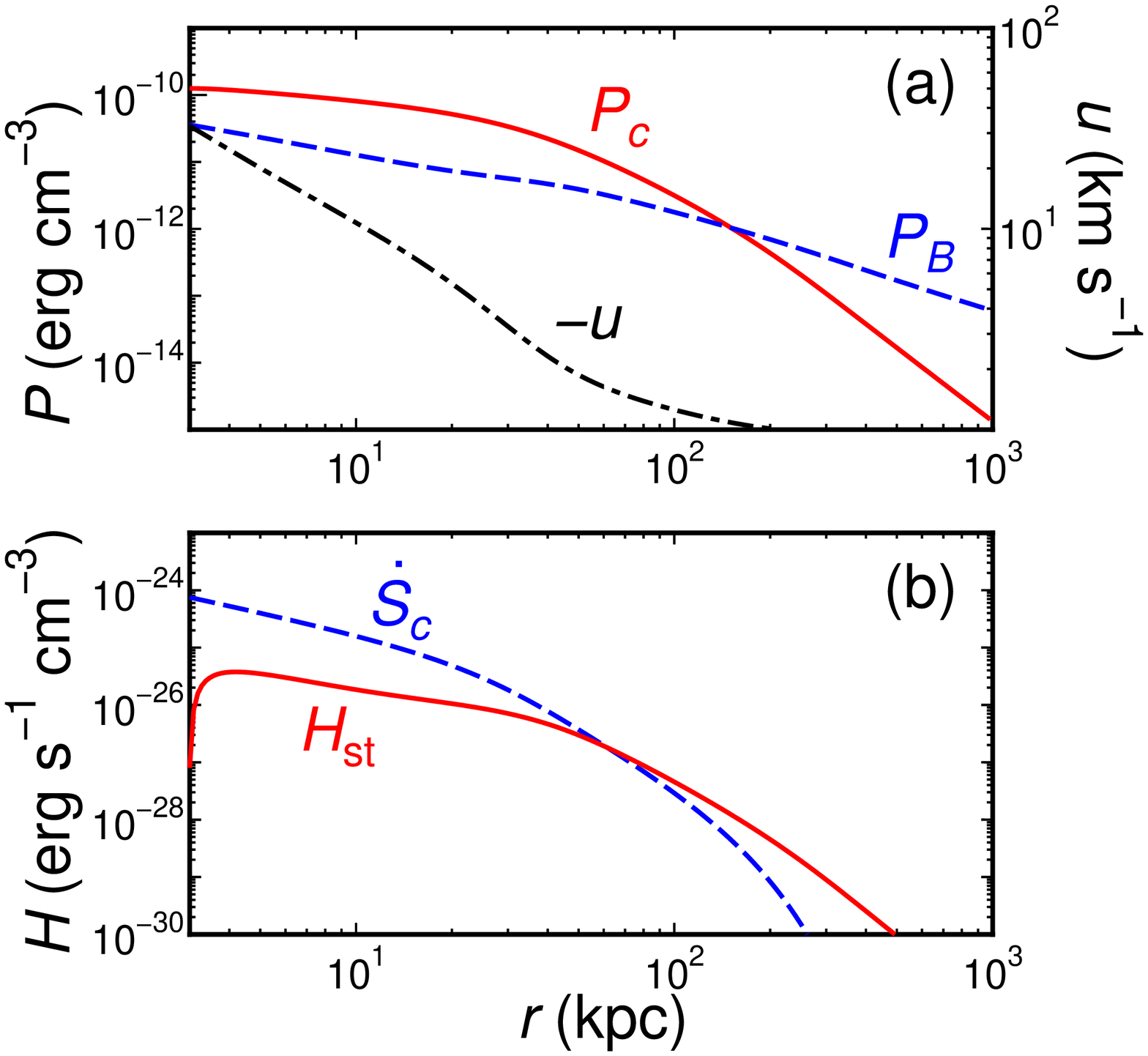} \caption{ Profiles of (a) CR
pressure (solid), magnetic pressure (dashed), and gas velocity
(dot-dashed), (b) heating rate of CR streaming (solid) and CR injection
rate (dashed) for A1795.}  \label{fig:H_a1795_vA}
\end{figure}

Fig.~\ref{fig:H_a1795_vA} shows the profiles of $P_c$, $P_B$, $u$,
$H_{\rm st}$, and $\dot{S}_c$ for the steady state solution of
A1795. The results are qualitatively the same for the other
clusters. The pressures ($P_c$ and $P_B$) and the infall velocity ($-u$)
increase toward the cluster centre (Fig.~\ref{fig:H_a1795_vA}a). Both
$P_c$ and $P_B$ are smaller than $P_g$ in the whole region. Although
$P_B>P_c$ for $r\ga 200$~kpc, $P_B$ is dynamically unimportant
there. Fig.~\ref{fig:H_a1795_vA}b shows that the slope of $H_{\rm st}$
is more gentle than that of $\dot{S}_c$ at $r\ga 4$~kpc. This reflects
that some of the CRs injected in the inner core region stream in the ICM
and heat the ICM in the outer core region.

\begin{table*}
 \centering
 \begin{minipage}{140mm}
  \caption{Cluster Parameters.\label{tab:cl}}
  \begin{tabular}{@{}cccccccc@{}}
  \hline
Cluster & $z$ & $v_{\rm st}$ & $T_{\rm in}$ & $T_{\rm out}$ & $n_0$ & 
$P_{c0}/P_{g0}$ & $\dot{M}$  \\
  & & & (keV) & (keV) & ($\rm cm^{-3}$)  & & ($M_\odot {\rm yr^{-1}}$) \\
 \hline
A1795 & 0.0632  & $v_A$ & 1.1 & 7.5 & 0.25  & 0.15 & -27.1 \\
      &         & $c_s$ & 1.5 & 7.5 & 0.17  & 0.03 & -19.7 \\
A2199 & 0.0309  & $v_A$ & 1.3 & 5   & 0.1   & 0.10 & -9.2  \\
      &         & $c_s$ & 1.5 & 5   & 0.1   & 0.03 & -10.4 \\
A2052 & 0.03549 & $v_A$ & 1.2 & 4.5 & 0.1   & 0.10 & -8.0  \\
      &         & $c_s$ & 1.3 & 4.5 & 0.1   & 0.03 & -7.4  \\
A2597 & 0.083   & $v_A$ & 1.3 & 4.5 & 0.1   & 0.20 & -10.4 \\
      &         & $c_s$ & 1.4 & 4.5 & 0.1   & 0.03 & -9.5  \\
\hline
\end{tabular}
\end{minipage}
\end{table*}

\subsection{Stability Analysis}
\label{sec:ana_vA}

\subsubsection{Lagrangian Perturbation Analysis}
\label{sec:lag_vA}

We investigate the stability of the steady state solutions obtained in
\S~\ref{sec:steady_vA}. In this subsection, we study the stability by a
Lagrangian perturbation analysis. We focus on the global thermal
instability in the ICM. The analysing method is based on that in
\citet{kim03a} and \citet{guo08b}.

The relation between a Lagrangian perturbation operator $\Delta$ and an
Eulerian perturbation $\delta$ is
\begin{equation}
\Delta = \delta + \mbox{\boldmath $\xi$}\cdot\nabla\:,
\end{equation}
where \mbox{\boldmath $\xi$} is the Lagrangian displacement of a fluid
element \citep[see][p. 127--147]{sha83a}. We consider only radial
perturbations and we define the radial component of \mbox{\boldmath
$\xi$} as $\xi=\Delta r$. In this case, the Lagrangian perturbation has
commutation relations of
\begin{equation}
\Delta\frac{d}{dt}=\frac{d}{dt}\Delta \:,
\end{equation}
\begin{equation}
\Delta\frac{\partial}{\partial r} = \frac{\partial}{\partial r}\Delta
-\frac{\partial\xi}{\partial r}\frac{\partial}{\partial r} \:
\end{equation}
\citep{sha83a}.

Equations~(\ref{eq:mom})--(\ref{eq:ec}) with $\kappa=D=H_{\rm
coll}=\Gamma_{\rm loss}=\partial P_B/\partial r=0$ can be rewritten as
\begin{equation}
\label{eq:mom3}
\rho\frac{du}{dt} = -\frac{\partial P_g}{\partial r} 
+ \rho g\:,
\end{equation}
\begin{equation}
\label{eq:eg3}
\frac{1}{\gamma_g -1}\frac{d P_g}{dt} - \frac{\gamma_g}{\gamma_g -1}
\frac{P_g}{\rho}\frac{d\rho}{dt}
=H_{\rm st} - \rho{\cal L}\:,
\end{equation}
\begin{equation}
\label{eq:ec3}
\frac{1}{\gamma_c -1}\frac{d P_c}{dt} - \frac{\gamma_c}{\gamma_c -1}
\frac{P_c}{\rho}\frac{d\rho}{dt}
+ \frac{\gamma_c}{\gamma_c -1}P_c
(\nabla\cdot\mbox{\boldmath $v$}_{\rm st})
=\dot{S}_c \:,
\end{equation}
where $d/dt$ is the Lagrangian time derivative, and $\mbox{\boldmath
$v$}_{\rm st}$ is the streaming velocity including the
direction. Equation~(\ref{eq:cont}) gives the mass flow rate,
$\dot{M}=4\pi r^2\rho u$.

We linearize equations~(\ref{eq:mom3})--(\ref{eq:ec3}). From these
equations, and useful relations with gas density and pressure
\begin{equation}
\label{eq:Drho}
\Delta\rho = -\rho\nabla\cdot\mbox{\boldmath $\xi$} \:,
\end{equation}
\begin{equation}
\Delta P_g = P_g\frac{\Delta T}{T} 
- P_g\nabla\cdot\mbox{\boldmath $\xi$}\:
\end{equation}
\citep{sha83a,guo08b}, we obtain following equations:
\begin{eqnarray}
\label{eq:mom4}
\lefteqn{\frac{d^2\xi}{dt^2} = 
\frac{P_g}{\rho}\frac{\partial}{\partial r}
(\nabla\cdot\mbox{\boldmath $\xi$})} \nonumber\\
&& - \frac{1}{\rho}\frac{\partial}{\partial r}
\left(P\frac{\Delta T}{T}\right) 
+ \frac{1}{\rho}\frac{\partial P_g}{\partial r}
\frac{\partial\xi}{\partial r}
-\xi\frac{d^2\xi}{dr^2} \:,
\end{eqnarray}
\begin{eqnarray}
\label{eq:eg4}
\lefteqn{\left(\frac{P_g}{\gamma_g-1}\frac{d}{dt} 
+ \rho T {\cal L}_T 
+ \frac{1}{\gamma_g-1}\frac{dP_g}{dt}
- \frac{\gamma_g}{\gamma_g-1}\frac{P_g}{\rho}\frac{d\rho}{dt}
\right) 
\frac{\Delta T}{T}} \nonumber \\
&&+ \left(P\frac{d}{dt} - \rho^2{\cal L}_\rho - H_{\rm st}\right)
(\nabla\cdot\mbox{\boldmath $\xi$}) - \Delta H_{\rm st} 
=0 \:,
\end{eqnarray}
\begin{equation}
\label{eq:ec4}
\frac{d}{dt}\left(\frac{\Delta P_c}{P_c} 
+ \gamma_c\nabla\cdot\mbox{\boldmath $\xi$}\right)
+ \gamma_c(\nabla\cdot\mbox{\boldmath $v$}_{\rm st})
= \frac{\Delta \dot{S}_c}{P_c}
- \frac{\dot{S}_c \Delta P_c}{P_c^2} \:,
\end{equation}
where ${\cal L}_T=\partial {\cal L}/\partial T|_\rho$ and ${\cal
L}_\rho=\partial {\cal L}/\partial \rho|_T$. 

Since $v_{\rm st}=v_A=B/\sqrt{4\pi\rho}\propto \rho^{\alpha}$, where
$\alpha=b-0.5$, perturbations of $v_{\rm st}$ can be
represented by
\begin{equation}
\label{eq:Dv_st2}
\Delta v_{\rm st} = \alpha v_{\rm st} \frac{\Delta \rho}{\rho}\:,
\end{equation}
\begin{eqnarray}
\label{eq:DDv_st2}
\lefteqn{\Delta(\nabla\cdot\mbox{\boldmath $v$}_{\rm st})
= -v_{\rm st}\left(\frac{2\xi}{r^2}
+ \frac{\alpha}{\rho}\frac{\partial\xi}{\partial r}
  \frac{\partial \rho}{\partial r}\right)}
 \nonumber\\
&&+\: \alpha v_{\rm st}\frac{\Delta \rho}{\rho}
\left(\frac{2}{r} + \frac{\alpha}{\rho}
\frac{\partial \rho}{\partial r}\right)
+ \alpha v_{\rm st}\frac{\partial}{\partial r}
\left(\frac{\Delta \rho}{\rho}\right) \:,
\end{eqnarray}
\begin{eqnarray}
\label{eq:DH_st2}
\lefteqn{\Delta H_{\rm st} = 
-v_{\rm st}\left(\alpha\frac{\Delta \rho}{\rho}
- \frac{\partial\xi}{\partial r}\right)\frac{\partial P_c}{\partial r}}
\nonumber\\
&&- v_{\rm st}\left[P_c\frac{\partial}{\partial r}
\left(\frac{\Delta P_c}{P_c}\right)
+ \frac{\Delta P_c}{P_c}\frac{\partial P_c}{\partial r}
\right] \:,
\end{eqnarray}
respectively. In the above equations, the density perturbation can be
replaced by $\Delta\rho/\rho=-\nabla\cdot\mbox{\boldmath $\xi$}$
(eq.~[\ref{eq:Drho}]).

Equations~(\ref{eq:dotSc}) and (\ref{eq:LAGN}) show that the source of
CRs can be rewritten as $\dot{S}_c=Q(r)\dot{M}$. We can write the
perturbation of the source as
\begin{equation}
\label{eq:DdotSc}
\Delta\dot{S}_c=\left(\frac{\Delta Q}{Q}
+ \frac{\Delta\dot{M}_{\rm in}}{\dot{M}}\right)\dot{S}_c \:,
\end{equation}
where $\Delta\dot{M}_{\rm in}$ is the perturbation of the mass flow rate
at $r=r_{\rm in}$. The perturbations in equation~(\ref{eq:DdotSc}) are
\begin{equation}
\Delta Q = \frac{dQ}{dr}\xi \:,
\end{equation}
\begin{equation}
\Delta\dot{M}_{\rm in}=\frac{\dot{M}}{u_{\rm
 in}}\left.\frac{\partial\xi}{\partial t}\right|_{r_{\rm in}}\:,
\end{equation}
where $u_{\rm in}=u(r_{\rm in})$.

We take $\xi$, $\Delta T$, and $\Delta P_c$ as independent variables. We
seek linear eigenmodes that behave as $\sim e^{\sigma t}$ with
time. Equations~(\ref{eq:mom4})--(\ref{eq:ec4}) may be rewritten as
\begin{eqnarray}
\label{eq:mom5}
\lefteqn{\left(\frac{P_g}{\rho}-u^2\right)\frac{d}{dr}
(\nabla\cdot\mbox{\boldmath $\xi$})
=\left(r\sigma^2 - r\frac{dg}{dr}\right)\frac{\xi}{r}
+ \frac{1}{\rho}\frac{d}{dr}\left(P_g\frac{\Delta T}{T}\right)}
\nonumber \\
&&- 2u^2\frac{d}{dr}\left(\frac{\xi}{r}\right)
+ \left(2\sigma u + u\frac{du}{dr} - \frac{1}{\rho}\frac{dP_g}{dr}\right)
\frac{d\xi}{dr} \:,
\end{eqnarray}
\begin{eqnarray}
\label{eq:eg5}
 \lefteqn{\left(\frac{P_g\sigma}{\gamma_g-1} + \rho T {\cal L}_T
+ \frac{u}{\gamma_g-1}\frac{dP_g}{dr}
- \frac{\gamma_g u}{\gamma_g-1}\frac{P_g}{\rho}\frac{d\rho}{dr}
\right)\frac{\Delta T}{T}} \nonumber\\
&&+ (P_g\sigma - \rho^2{\cal L}_\rho - H_{\rm st})
(\nabla\cdot\mbox{\boldmath $\xi$}) - \Delta H_{\rm st} \nonumber\\
&&+ P_gu\frac{d}{dr}(\nabla\cdot\mbox{\boldmath $\xi$})
+ \frac{P_gu}{\gamma_g-1}\frac{d}{dr}\left(\frac{\Delta T}{T}\right) = 0
\:,
\end{eqnarray}
\begin{eqnarray}
\label{eq:ec5}
\lefteqn{\sigma\frac{\Delta P_c}{P_c} 
+ (v_{\rm st}+u)\frac{d}{dr}\left(\frac{\Delta P_c}{P_c}\right)
+ \sigma\gamma_c(\nabla\cdot\mbox{\boldmath $\xi$}) 
+ \gamma_c u \frac{\partial}{\partial r}
(\nabla\cdot\mbox{\boldmath $\xi$})} \nonumber\\
&&+ \gamma_c \Delta(\nabla\cdot\mbox{\boldmath $v$}_{\rm st})
= \frac{\Delta \dot{S}_c}{P_c}
- \frac{\dot{S}_c \Delta P_c}{P_c^2} \:.
\end{eqnarray}
We omit the dependence of $e^{\sigma t}$ hereafter.
Equations~(\ref{eq:mom5})--(\ref{eq:ec5}) are first-order differential
equations for the four variables $\xi/r$, $\Delta T/T$, $\Delta
P_c/P_c$, and $d(\xi/r)/dr$. We numerically solve these equations as an
eigenvalue problem, where the eigenvalue is the growth rate $\sigma$
\citep{kim03a,guo08b}, using {\it Mathematica~9}. Following the previous
studies, we set five boundary conditions. At the inner boundary
($r=r_{\rm in}$), we give three conditions:
\begin{equation}
\label{eq:bcin1}
\xi/r = 1 \:,
\end{equation} 
\begin{equation}
\label{eq:bcin2}
\frac{d}{dr}\left(\frac{\xi}{r}\right) = 0 \:,
\end{equation}
\begin{eqnarray}
\label{eq:bcin3}
\lefteqn{\Delta (r^2 \tilde{u} P_c)
 = 2(u+v_{\rm st})P_c r \xi} \nonumber \\
&& +\left(\sigma\xi + u\frac{\partial \xi}{\partial r}
+ \Delta v_{\rm st}\right)P_c r^2 
 + r^2 (u+v_{\rm st})\Delta P_c \nonumber \\
& &= 0
\:.
\end{eqnarray}
Equation~(\ref{eq:bcin1}) is a normalisation condition, and
equation~(\ref{eq:bcin2}) guarantees the regularity of the
solutions. Equation~(\ref{eq:bcin3}) means that the perturbed CR flux is
zero at the cluster centre. At the outer boundary ($r=r_{\rm out}$), we
adopt the two conditions:
\begin{equation}
\label{eq:bcout1}
\xi = 0\:,
\end{equation}
\begin{equation}
\label{eq:bcout2}
\Delta T = 0\:,
\end{equation}
because the cooling time of the ICM is much larger than the cluster age.

The background ICM and CR profiles are given by the steady state
solutions derived in \S~\ref{sec:steady_vA}. We search $\sigma$ that
satisfies equations~(\ref{eq:mom5})--(\ref{eq:ec5}) and boundary
conditions (\ref{eq:bcin1})--(\ref{eq:bcout2}) in the range of
$(10^4\:{\rm Gyr})^{-1}<\sigma<(10^{-4}\:{\rm Gyr})^{-1}$.  Since the
equations are rather complicated, it takes a long time to find
solutions. Thus, we limited our search to real $\sigma$ and did not
search imaginary $\sigma$. Following \citet{kim03a} and~\citet{guo08b},
we first fix $\sigma$ and give arbitrarily $\Delta T/T$ at $r=r_{\rm
in}$. We then integrate equations~(\ref{eq:mom5})--(\ref{eq:ec5}) from
$r=r_{\rm in}$ to $r=r_{\rm out}$ and check whether the first outer
boundary condition~(\ref{eq:bcout1}) are satisfied. If not, we try
another $\Delta T$. We use the second outer boundary
condition~(\ref{eq:bcout2}) as a discriminant for solutions.

\begin{figure}
\includegraphics[width=84mm]{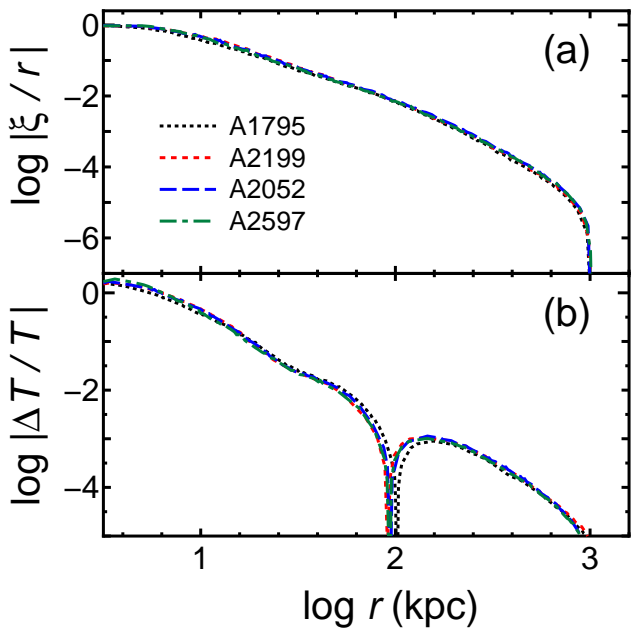} \caption{Eigenfunctions of the
stable mode for A1795 (dotted), A2199 (short-dashed), A2057
(long-dashed), and A2597 (dot-dashed) plotted as functions of
radius. (a) $\xi/r$, (b) $\Delta T/T$. The streaming velocity is $v_{\rm
st}=v_A$.} \label{fig:eig_vA}
\end{figure}

As a result, we could not find any solutions with positive $\sigma$ in
the above range for the four clusters. This means that perturbations do
not grow and the steady state solutions are quite stable even if we do
not include thermal conduction, which has often been used to stabilise
heating. On the other hand, we found decaying modes ($\sigma<0$).
Fig.~\ref{fig:eig_vA} shows the eigenfunctions for the lowest order
mode. For the mode shown in the figure, $-\sigma\sim 0.3\:\rm Gyr^{-1}$
for the four clusters.

\subsubsection{Numerical Simulations}
\label{sec:num_vA}

In the analysis in \S~\ref{sec:lag_vA}, we did not perform a complete
parameter search; we did not examine imaginary $\sigma$ and local
instabilities, for example.  Thus, in this subsection, we
supplementarily study the stability of the steady state solutions using
numerical simulations. We solve equations~(\ref{eq:cont})--(\ref{eq:ec})
with $\kappa=D=H_{\rm coll}=\Gamma_{\rm loss}=\partial P_B/\partial
r=0$. The hydrodynamic part of the equations is solved by a second-order
advection upstream splitting method (AUSM) based on \citet{lio93a}. We
use 300 unequally spaced meshes in the radial coordinate to cover a
region with a radius of $r_{\rm out}=1$~Mpc. The inner boundary is set
at $r_{\rm in}=3$~kpc. The innermost mesh has a width of 90~pc, and the
width of the outermost mesh is 17~kpc. While variables have zero
gradients at the inner boundary, density and pressure are fixed at their
initial values at the outer boundary. We use the steady state solutions
obtained in \S~\ref{sec:steady_vA} as initial conditions ($t=0$).

Figs.~\ref{fig:Tn_a1795_vA}-\ref{fig:Tn_a2597_vA} show the results of
the calculations.  For all the four clusters, the ICM is stably heated
at least within the age of the Universe ($t\sim 14$~Gyr). For A2199,
A2057, and A2597, profiles are almost identical to the initial ones at
$t\la 40$~Gyr. However, local instabilities start developing at $t\ga
40$~Gyr at $r\sim 50$~kpc and it later affects the inner profiles. In
fact, it has been indicated that the heating via CR streaming is not
locally stable \citep{loe91,pfr13a}. For A1795, the local instabilities
do not develop until $t=100$~Gyr.

Although the local instabilities develop for the three clusters, they
are not very radical. We found that our assumption of smaller $b$
contributes to the suppression of rapid development of the local
instabilities. Since we assumed that $b=0.4$, the Alfv\'en velocity is
$v_A=B/\sqrt{4\pi\rho}\propto \rho^{-0.1}$. If excessive cooling
increases $\rho$ at the cluster centre for example, it reduces $v_A$
there. The smaller $v_A$ prevents CRs from escaping from the cluster
centre and increases $P_c$ and $|\partial P_c/\partial r|$ around the
centre. Since $H_{\rm st}\propto |\partial P_c/\partial r|$, the cluster
centre is well-heated. The same mechanism should work when $v_{\rm
st}=c_s$, because excessive cooling reduces the streaming
velocity. Using numerical simulations, we found that the local
instabilities we discussed here are suppressed by moderate thermal
conduction of the level of $\sim 1$\% of the Spitzer conductivity.

\section{Discussion}
\label{sec:dis}

\subsection{Evolution with time}

Because of possible cluster mergers and the change of AGN activities, it
is likely that the ICM profiles sometimes deviate from the steady state
solutions we studied above. Thus, we study the time scale in which the
perturbed ICM profiles return to the steady ones.

As an example, we choose the ICM profile of the steady state solution
for A1795 when $v_{\rm st}=v_A$ as a fiducial profile
(Table~\ref{tab:cl} and Fig.~\ref{fig:Tn_a1795_vA}). In this model, the
efficiency $\epsilon$ in equation~(\ref{eq:LAGN}) was
$\epsilon=2.5\times 10^{-4}$.  We change the efficiency and construct
new steady state solutions, while fixing the boundary parameters ($n_0$,
$T_{\rm in}$, $T_{\rm out}$, and $P_{c0}$). We obtain $\dot{M}=19.0\:
M_\odot\rm\: yr^{-1}$ and $39.4\: M_\odot\rm\: yr^{-1}$ for
$\epsilon=5\times 10^{-4}$ and $1\times 10^{-4}$, respectively. Their
temperature and density profiles are not much different from those in
Fig.~\ref{fig:Tn_a1795_vA}, because the boundary conditions are the
same. Using those profiles as the initial ones, we numerically solve
equations~(\ref{eq:cont})--(\ref{eq:ec}) with $\kappa=D=H_{\rm
coll}=\Gamma_{\rm loss}=\partial P_B/\partial r=0$ for $v_{\rm st}=v_A$
and $\epsilon=2.5\times 10^{-4}$. This may be the situation where the
heating efficiency of the central AGN suddenly changes at $t=0$.

\begin{figure}
\includegraphics[width=84mm]{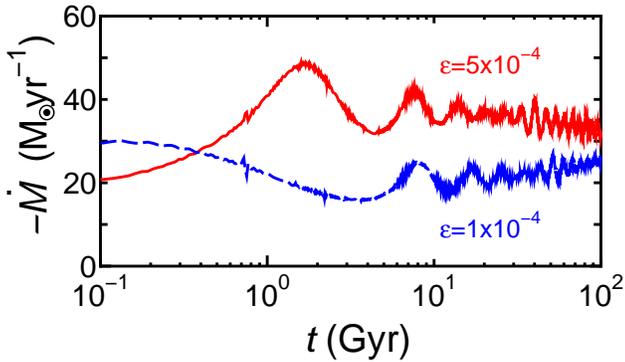} \caption{Evolution of $\dot{M}$
 for $\epsilon=5\times 10^{-4}$ (solid) and $1\times 10^{-4}$ (dashed)
 at $t=0$.} 
\label{fig:evo}
\end{figure}

In Fig.~\ref{fig:evo}, we show the evolution of $\dot{M}$, which is
proportional to $L_{\rm AGN}$. The mass flow rates oscillate on a time
scale of $\sim 3$~Gyr, which is comparable to the time scale of the
decaying mode ($-1/\sigma$) that we studied in \S~\ref{sec:lag_vA}. The
amplitude of the oscillation gradually decreases and $\dot{M}$ converges
to that of the fiducial model of $\epsilon=2.5\times 10^{-4}$
($\dot{M}=-27.1\: M_\odot\rm\: yr^{-1}$). Since the time scale of the
first large oscillation ($\sim 3$~Gyr) is smaller than the cluster
lifetime ($\sim$10~Gyr), we think that the discussions based on the
steady state solutions are justified.

\subsection{Radio mini-halos}

In cool cores in some clusters, diffuse radio emissions called
mini-halos have been observed. Our model predicts that the diffuse
emissions originate from the CR protons heating the cluster cores
(Paper~III). In particular, the observed radio profiles can be nicely
reproduced by our model. Recently, it has also been proposed that the CR
electrons responsible for the radio emission have been accelerated by
turbulence generated by the sloshing of the cool core gas
\citep[e.g.][]{maz08a,zuh13a}. Contrary to these models, our model does
not require the sloshing because the CR protons are accelerated by the
central AGNs.

While there is no direct relationship between the mini-halos and the
sloshing in our model, an apparent relationship could be observed. In
our model, mini-halos are associated with the AGNs that are active in
cool cores. Meanwhile, structures related to the sloshing (e.g. cold
fronts) are often produced around the cool cores
\citep{fuj04,asc06}. Moreover, in our model, the CR electrons
responsible for the radio emission are created through the interaction
between the CR protons and the target gas protons. Thus, more electrons
are created in denser gas, and brighter radio emission would be observed
in denser cores or on the denser side of a cold front associated with
the sloshing. Of course, our model predicts that even clusters without
sloshing can have the mini-halos. In the sloshing model, the morphology
of the simulated mini-halos is complicated and depends on observing
frequencies, because the sloshing is temporary and the CR acceleration
occurs in the region where turbulence is developing
\citep{zuh13a}. Thus, the radio emission could be strong even not at the
cluster centre. This is not generally true in our model. because the
radio emission tends to be strong at the cluster centre where the gas
and CR densities are the highest.

On the other hand, we do not think that our model can explain Mpc-scale
radio halos that are often observed in merging clusters without cool
cores, because AGNs are active in cool cores, and dense gas in cool
cores is required to produce the CR electrons. Those Mpc-scale radio
halos may be produced by electrons accelerated in turbulence in the ICM
\citep{bru01,pet01b,ohn02,fuj03,bru11}.

\section{Conclusions}
\label{sec:con}

We have investigated the stability of CR heating in clusters. The CRs
are accelerated at the central AGNs, and heat the ICM through CR
streaming. First, we obtained steady state solutions of the ICM and CR
profiles so that they are consistent with observations. The solutions
are obtained when the magnetic fields are strong enough and their
association with the ICM is relatively weak. Then, we analysed the
stability of the solutions analytically and numerically. For the
analytic approach, we adopted a Lagrangian stability analysis and found
that there are no globally unstable modes. For the numerical approach,
we followed the evolution of the solutions for 100~Gyr and confirmed
that the solutions are quite stable. These results, as well as the
consistency with radio observations (Paper~III), make the CR heating an
attractive solution of the cooling flow problem.

\section*{Acknowledgments}

We appreciate the referee's useful comments. We thank F.~Takahara,
T.~Tsuribe, D.~Nagai, and L.~Rudnick for useful discussion.  This work
was supported by KAKENHI (Y.~F.: 23540308, Y.~O.: 24.8344).

\appendix

\section[]{Sound Velocity Case}
\label{sec:cs}

Here, we consider the case where the streaming velocity is the sound
velocity of the ICM ($v_{\rm st}=c_{\rm s}$). We set $P_B=0$ because
$v_{\rm st}$ does not depend on magnetic fields.

\subsection{Steady State Solutions}
\label{sec:steady_cs}

The steady state solutions can be obtained by replacing $v_A$ by $c_s$
in \S~\ref{sec:steady_vA}. We solve
equations~(\ref{eq:mom2})--(\ref{eq:ec2}) with boundary
conditions~(\ref{eq:bc_n0})--(\ref{eq:bc_Pc0}). In order to match the
solutions with observations, we adjust $n_0$, $T_{\rm in}$, and $P_{\rm
c0}$, while $T_{\rm out}$ is uncharged from that in
\S~\ref{sec:steady_vA}.  We found that $P_{c0}/P_{g0}\sim 0.01$--0.1
from comparison with radio observations (Fig.~14 in Paper~III). Thus, we
assume that $P_{c0}/P_{g0}=0.03$ for all the four clusters.

\begin{figure}
\includegraphics[width=84mm]{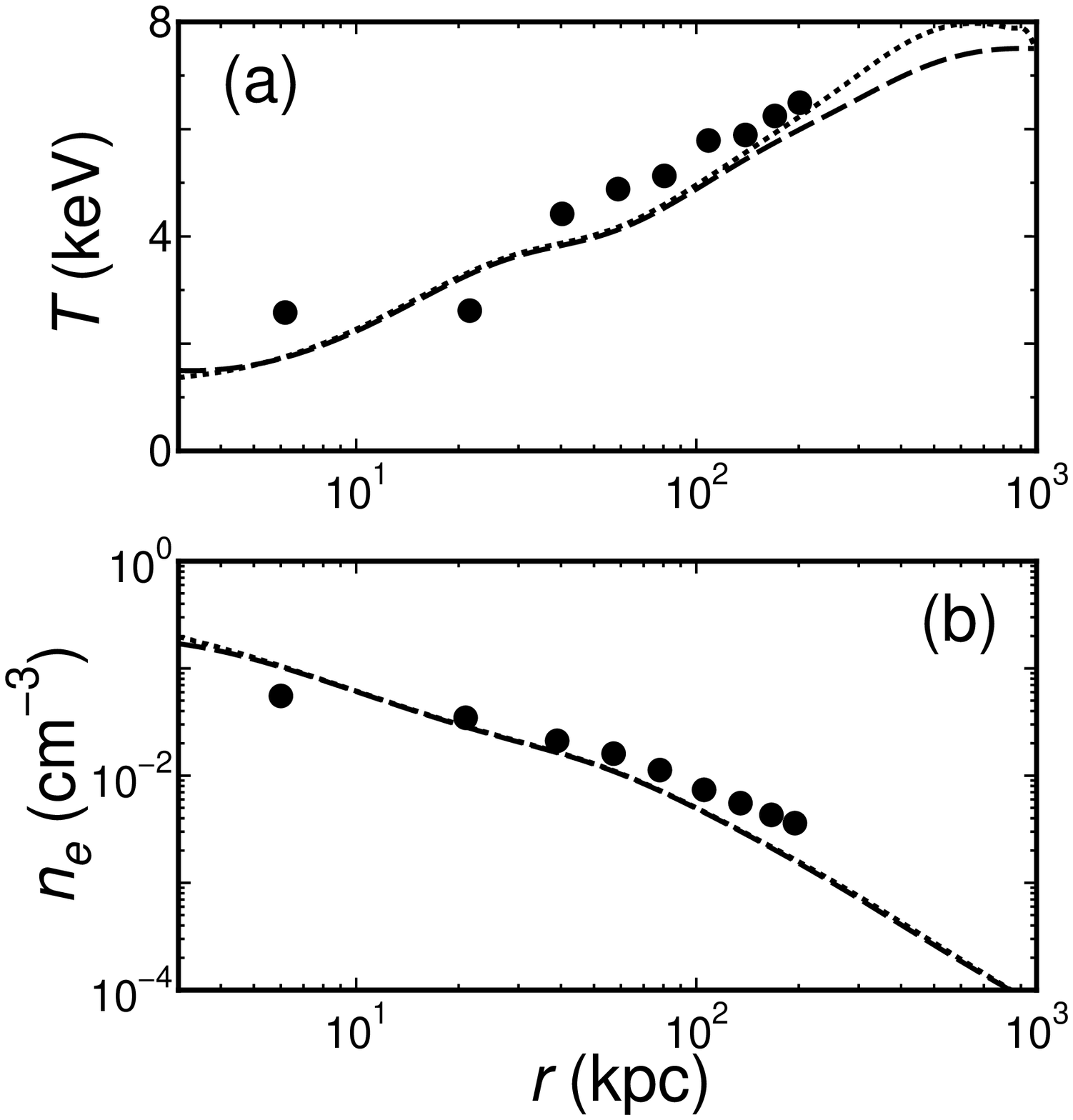} \caption{(a) Temperature and (b)
density profiles for A1795. Dashed lines show the steady state solution
or the initial profiles for the numerical simulation. Dotted lines are
the results of numerical simulation at $100$~Gyr. Filled circles
represent observations \citep{ett02a}. Error bars are omitted.}
\label{fig:Tn_a1795_cs}
\end{figure}

\begin{figure}
\includegraphics[width=84mm]{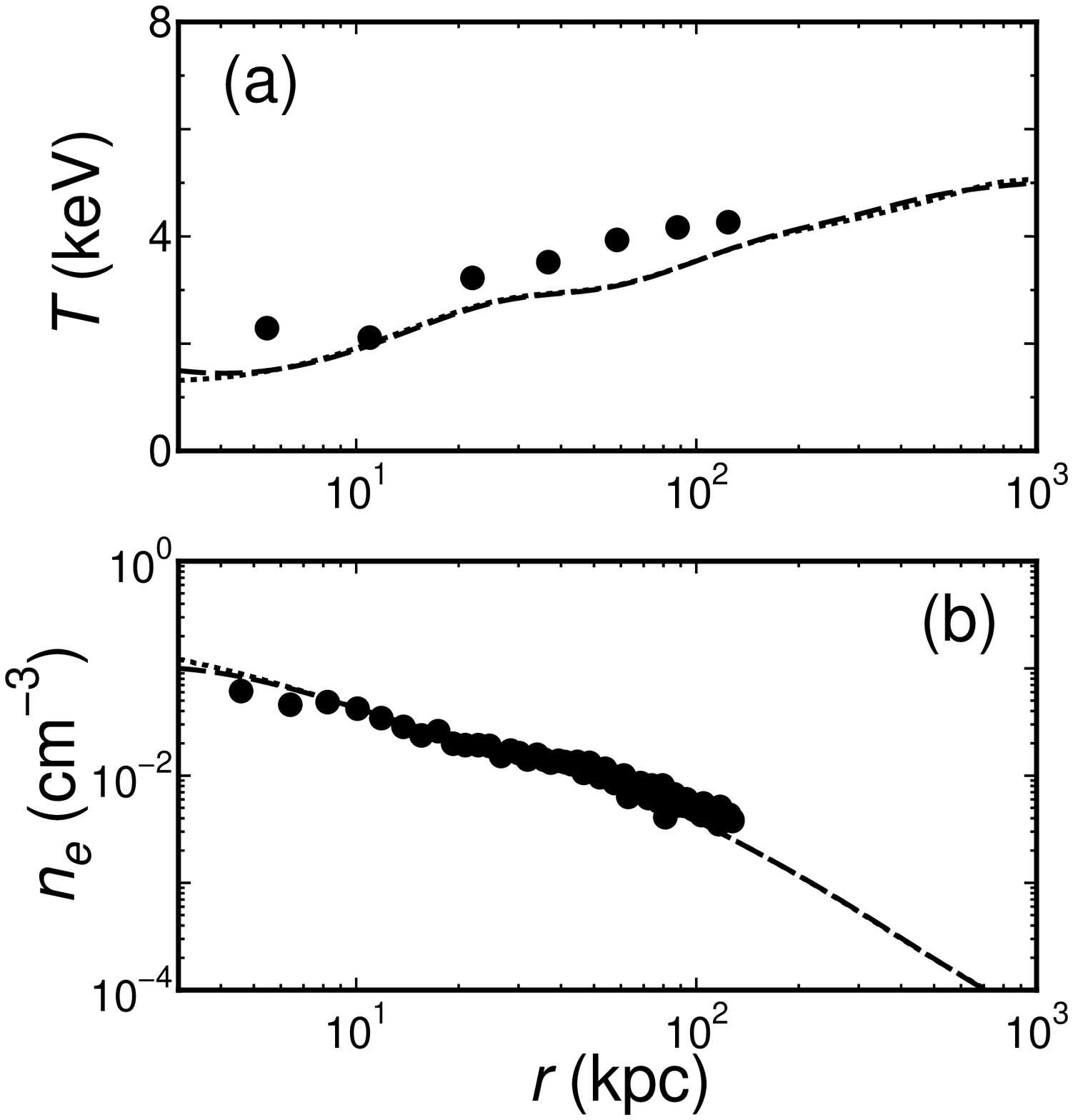} \caption{Same as
Fig.~\ref{fig:Tn_a1795_cs} but for A2199. Filled circles represent
observations \citep{jon02a}.}
\label{fig:Tn_a2199_cs}
\end{figure}

\begin{figure}
\includegraphics[width=84mm]{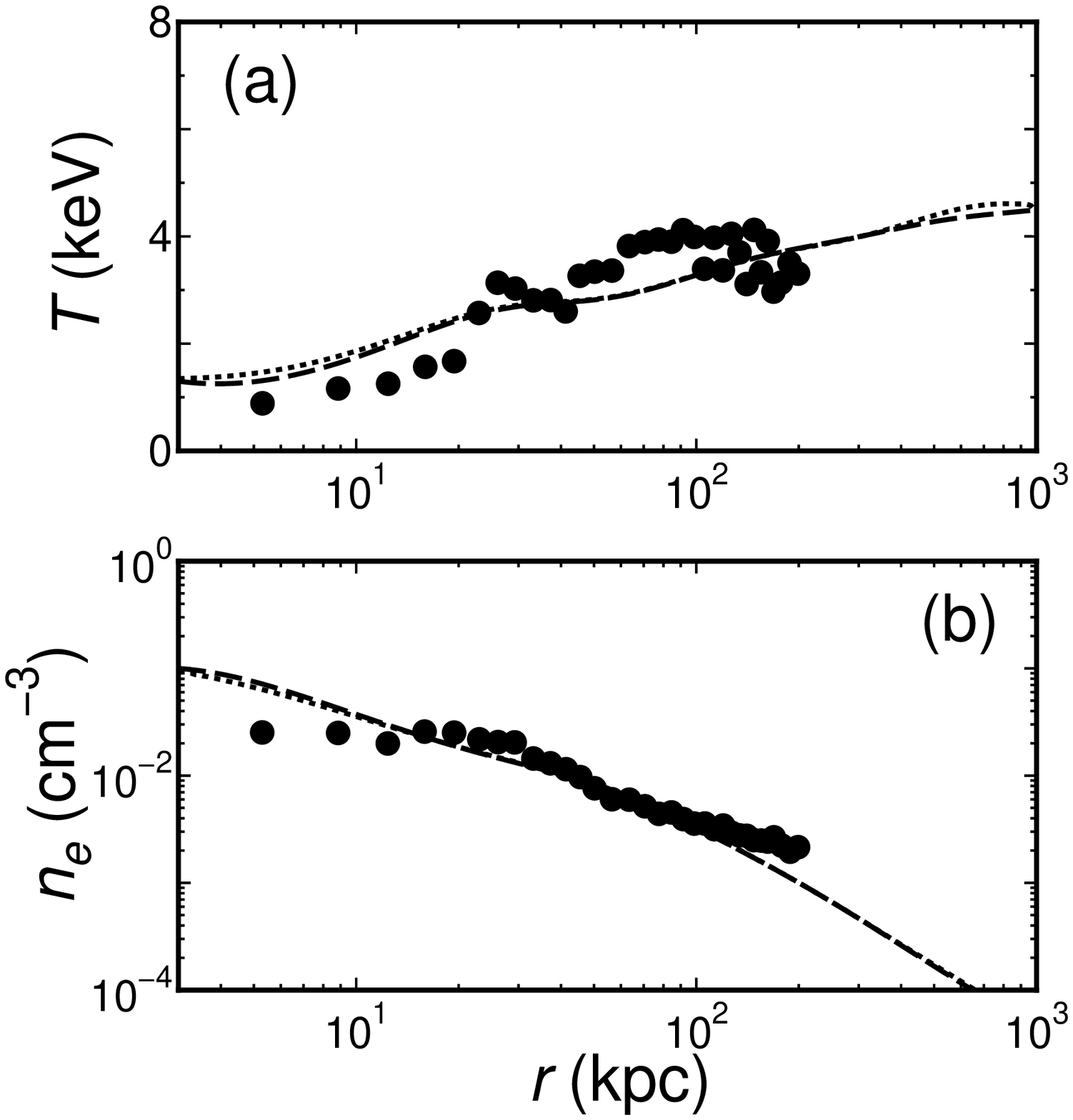} \caption{Same as
Fig.~\ref{fig:Tn_a1795_cs} but for A2052. Filled circles represent
observations \citep{bla11a}.}
\label{fig:Tn_a2052_cs}
\end{figure}

\begin{figure}
\includegraphics[width=84mm]{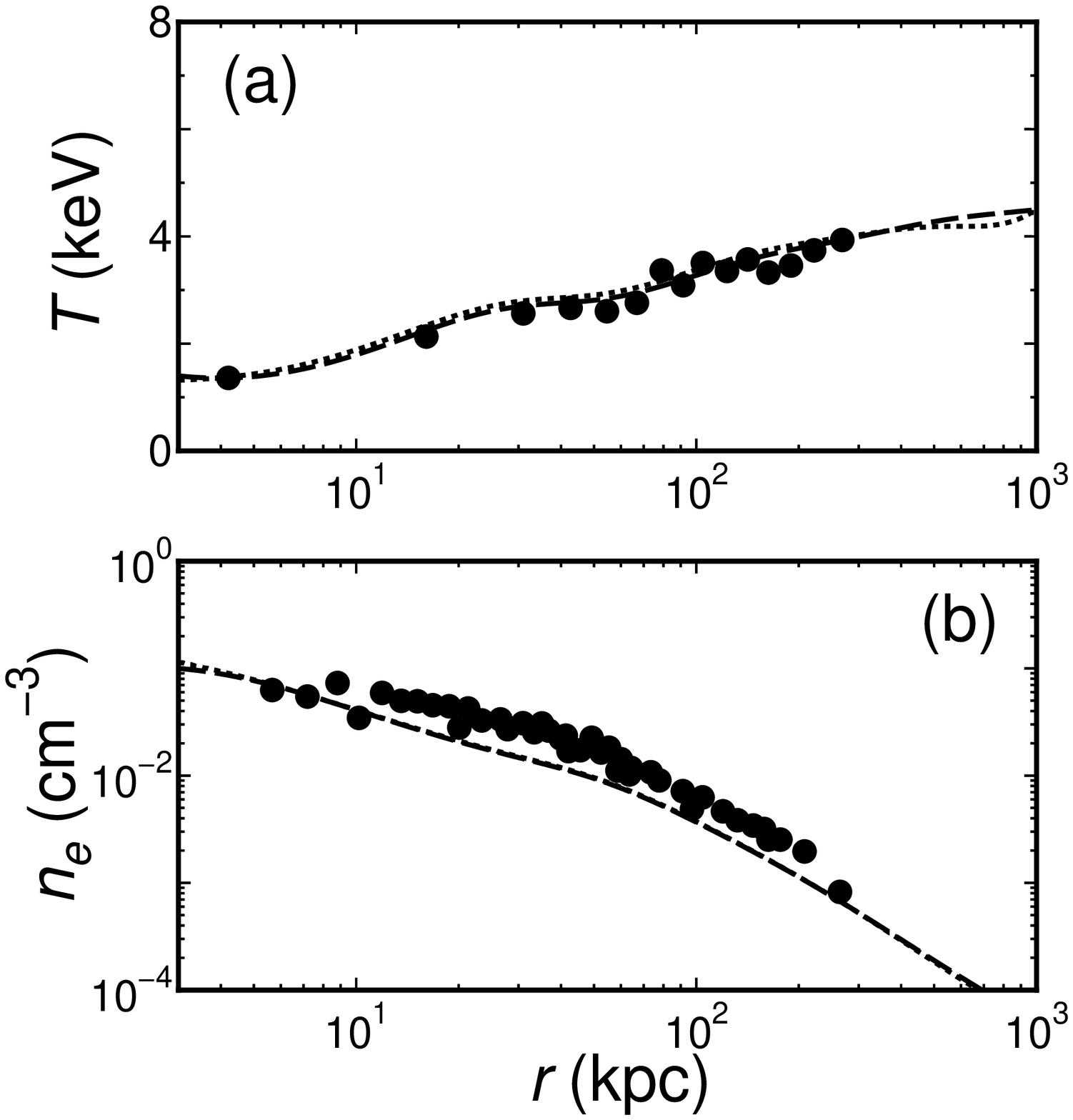} \caption{Same as
Fig.~\ref{fig:Tn_a1795_cs} but for A2597. Filled circles represent
observations \citep{mcn01a}.}
\label{fig:Tn_a2597_cs}
\end{figure}

\begin{figure}
\includegraphics[width=84mm]{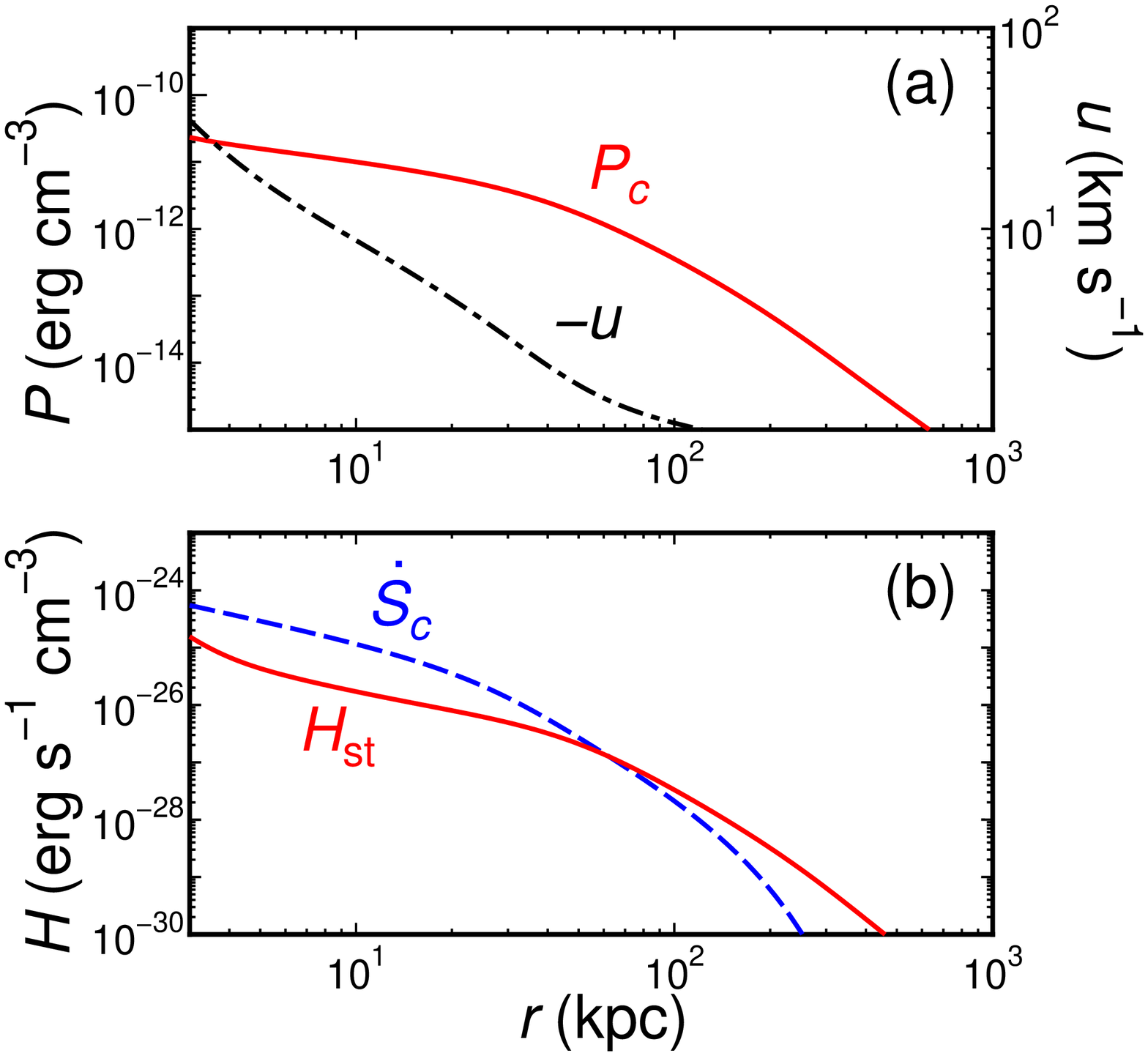} \caption{Profiles of (a) CR
pressure (solid) and gas velocity (dot-dashed), (b) heating rate of CR
streaming (solid) and CR injection rate (dashed) for A1795.}
\label{fig:H_a1795_cs}
\end{figure}

The dashed lines in Figs.~\ref{fig:Tn_a1795_cs}-\ref{fig:Tn_a2597_cs}
show the steady state solutions for the four clusters. The temperature
and density profiles generally reproduce the observations. The
parameters we adopted are shown in Table~\ref{tab:cl} ($v_{\rm
st}=c_s$).  Fig.~\ref{fig:H_a1795_cs} shows the profiles of $P_c$, $u$,
$H_{\rm st}$, and $\dot{S}_c$ for the steady state solution of
A1795. The results are qualitatively the same for the other
clusters. Compared to Fig.~\ref{fig:H_a1795_vA}a, $P_c$ is smaller
because $v_{\rm st}=c_s$ is larger than $v_A$, and it enhances the
escape of the CRs
(Fig.~\ref{fig:H_a1795_cs}a). Fig.~\ref{fig:H_a1795_cs}b shows that the
heating ($H_{\rm st}$) is more widely distributed than the CR injection
($\dot{S}_c$) as is the case of $v_{\rm st}=v_A$
(Fig.~\ref{fig:H_a1795_vA}b).

\subsection{Stability Analysis}
\label{sec:ana_cs}

\subsubsection{Lagrangian Perturbation Analysis}
\label{sec:lag_cs}

As we did in \S~\ref{sec:lag_vA}, we solve
equations~(\ref{eq:mom5})--(\ref{eq:ec5}) under the five boundary
conditions~(\ref{eq:bcin1})--(\ref{eq:bcout2}). However, the terms
including $v_{\rm st}$ (equations.~[\ref{eq:Dv_st2}]--[\ref{eq:DH_st2}])
must be modified.

Since we assumed that $v_{\rm st}=c_s\propto T^{1/2}$, perturbations of
$v_{\rm st}$ can be represented by
\begin{equation}
\label{eq:Dv_st}
\Delta v_{\rm st} = \frac{v_{\rm st}}{2}\frac{\Delta T}{T}\:,
\end{equation}
\begin{eqnarray}
\label{eq:DDv_st}
\lefteqn{\Delta(\nabla\cdot\mbox{\boldmath $v$}_{\rm st})
= -v_{\rm st}\left(\frac{2\xi}{r^2}
+ \frac{1}{2T}\frac{\partial\xi}{\partial r}
  \frac{\partial T}{\partial r}\right)}
 \nonumber\\
&&+ v_{\rm st}\frac{\Delta T}{T}
\left(\frac{1}{r} + \frac{1}{4T}\frac{\partial T}{\partial r}\right)
+ \frac{v_{\rm st}}{2}\frac{\partial}{\partial r}
\left(\frac{\Delta T}{T}\right) \:.
\end{eqnarray}
The perturbation of the heating by the CR streaming is
\begin{eqnarray}
\label{eq:DH_st}
\lefteqn{\Delta H_{\rm st} = -v_{\rm st}
\left(\frac{1}{2}\frac{\Delta T}{T}
- \frac{\partial\xi}{\partial r}\right)\frac{\partial P_c}{\partial r}}
\nonumber\\
&& - v_{\rm st}\left[P_c\frac{\partial}{\partial r}
\left(\frac{\Delta P_c}{P_c}\right)
+ \frac{\Delta P_c}{P_c}\frac{\partial P_c}{\partial r}
\right] \:.
\end{eqnarray}

\begin{figure}
\includegraphics[width=84mm]{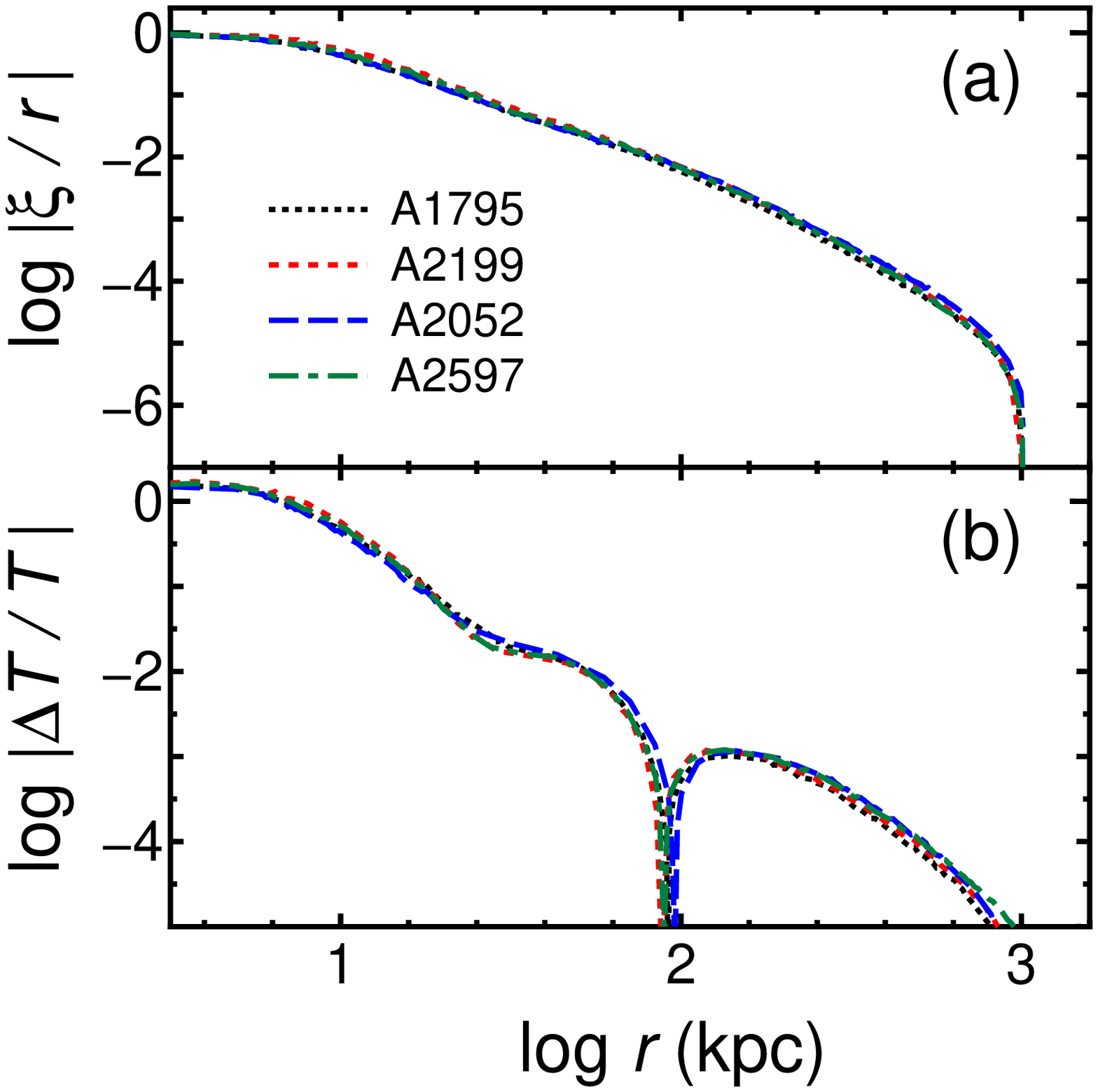} \caption{Same as
 Fig.~\ref{fig:eig_vA} but for $v_{\rm
st}=c_s$} \label{fig:eig_cs}
\end{figure}

The background ICM and CR profiles are given by the steady state
solutions derived in \S~\ref{sec:steady_cs}. We search $\sigma$ that
satisfies equations~(\ref{eq:mom5})--(\ref{eq:ec5}) and boundary
conditions (\ref{eq:bcin1})--(\ref{eq:bcout2}) in the range of
$(10^4\:{\rm Gyr})^{-1}<\sigma<(10^{-4}\:{\rm Gyr})^{-1}$. As is the
case of $v_{\rm st}=v_A$, we could not find any solutions with positive
$\sigma$ in the above range for the four clusters. On the other hand, we
found decaying modes ($\sigma<0$).  Fig.~\ref{fig:eig_cs} shows the
eigenfunctions for the lowest order mode. For the mode shown in the
figure, $-\sigma=0.4$, 0.4, 0.2, and $0.3\:\rm Gyr^{-1}$ for A1795,
A2199, A2057, and A2597, respectively.

\subsubsection{Numerical Simulations}
\label{sec:num_cs}

As we did in \S~\ref{sec:num_vA}, we numerically solve
equations~(\ref{eq:cont})--(\ref{eq:ec}) with $\kappa=D=H_{\rm
coll}=\Gamma_{\rm loss}=\partial P_B/\partial r=0$ for $v_{\rm
st}=c_s$. The initial ($t=0$) profiles are given by the steady state
solutions we obtained in \S~\ref{sec:steady_cs}.

Dotted lines in Figs.~\ref{fig:Tn_a1795_cs}--\ref{fig:Tn_a2597_cs} show
the profiles at $t=100$~Gyr. They are almost identical to those at $t=0$
(dashed lines). Contrary to the case of $v_{\rm st}=v_A$
(\S~\ref{sec:num_vA}), local instabilities do not develop for
$t<100$~Gyr for all the four clusters. The results indicate that the
steady solutions are very stable.

\end{document}